\documentclass[a4paper]{article}
\pdfoutput=1
\usepackage{icrc2013}
\newcommand{\V}[1]{\mathbf{#1}}

\title{On the Contribution of Pulsars to  the Positron Fraction in Cosmic Rays}
\shorttitle{Positron fraction interpretation}

\authors{S. Della Torre$^{1}$, M. Gervasi$^{1,2}$, P.G. Rancoita$^{1}$, D. Rozza$^{1,3,4}$, A. Treves$^{1,3}$.}
\afiliations
{
$^1$ \textit{INFN Sezione di Milano Bicocca, I-20126 Milano, Italy}\\
$^2$ \textit{Universit\`a di Milano Bicocca, I-20126 Milano, Italy} \\
$^3$ \textit{Universit\`a dell'Insubria, I-22100 Como, Italy}\\
$^4$ \textit{European Organization for Nuclear Research, CERN, CH-1211 Geneva 23, Switzerland}
\scriptsize{}
}
\email{Davide.Rozza@mib.infn.it}

\abstract{Several cosmic ray experiments have measured the positron fraction up to few hundred GeV. Their data have revealed an excess of positrons above 10 GeV that is not consistent with the secondary production of these particles in the interstellar medium. A primary source like dark matter or astrophysical sources (e.g pulsars and their nebulae) were considered to account for such an excess. In this paper we analyse the possibility of a primary positron production due to pulsars. Under the assumption of equal initial spectra at the source for positrons, electrons, and gamma-rays we study the propagation of particle spectra using a diffusion model in the Galaxy. We focused our analysis on the Vela and Crab  pulsars and their associated nebulae, which are well observed in gamma-rays. Comparison with experimental  data is reported. The propagated  positron and electron spectra generated from these sources result in a positron ratio, which is largely inconsistent with the excess observed by PAMELA and AMS.}


\begin{document}

\maketitle

\section{Introduction}
Cosmic ray positrons may be created in secondary production processes. These interactions of primary cosmic ray nuclei with the interstellar medium (ISM) produce pions and kaons that decay in positrons \cite{MS98}. The measure of their flux at 10 GeV is $\sim5-10\%$ the electron flux. Following this theory, the positron fraction (ratio between the positron and the sum of electron plus positron flux) is expected to decrease at high energy.\\
Extending the results of the previous experiments, AMS-02 \cite{AMS02} detected an increasing positron fraction between 10 GeV and 350 GeV. This excess has been interpreted in different ways. Positron-electron pairs can be produced from a star population e.g. pulsars \cite{Yin2013}, or to dark matter annihilation \cite{Ibe2013}. We concentrate our attention on the pulsar contribution.\\
High energy positrons are supposedly produced in pulsars as showers due to the photon - magnetic field interaction (see e.g. the seminal paper by Sturrock 1971 \cite{Sturrock71}). In the nebula, that can surround the pulsar, the interaction is with matter, and high energy electrons, positrons and photons may derive through the pion production. In both cases one can assume as a first approximation that particles and gamma rays have similar spectra at high energy.

\section{The injection spectrum}
\begin{figure}[!t]
  \centering
  \includegraphics[width=0.45\textwidth]{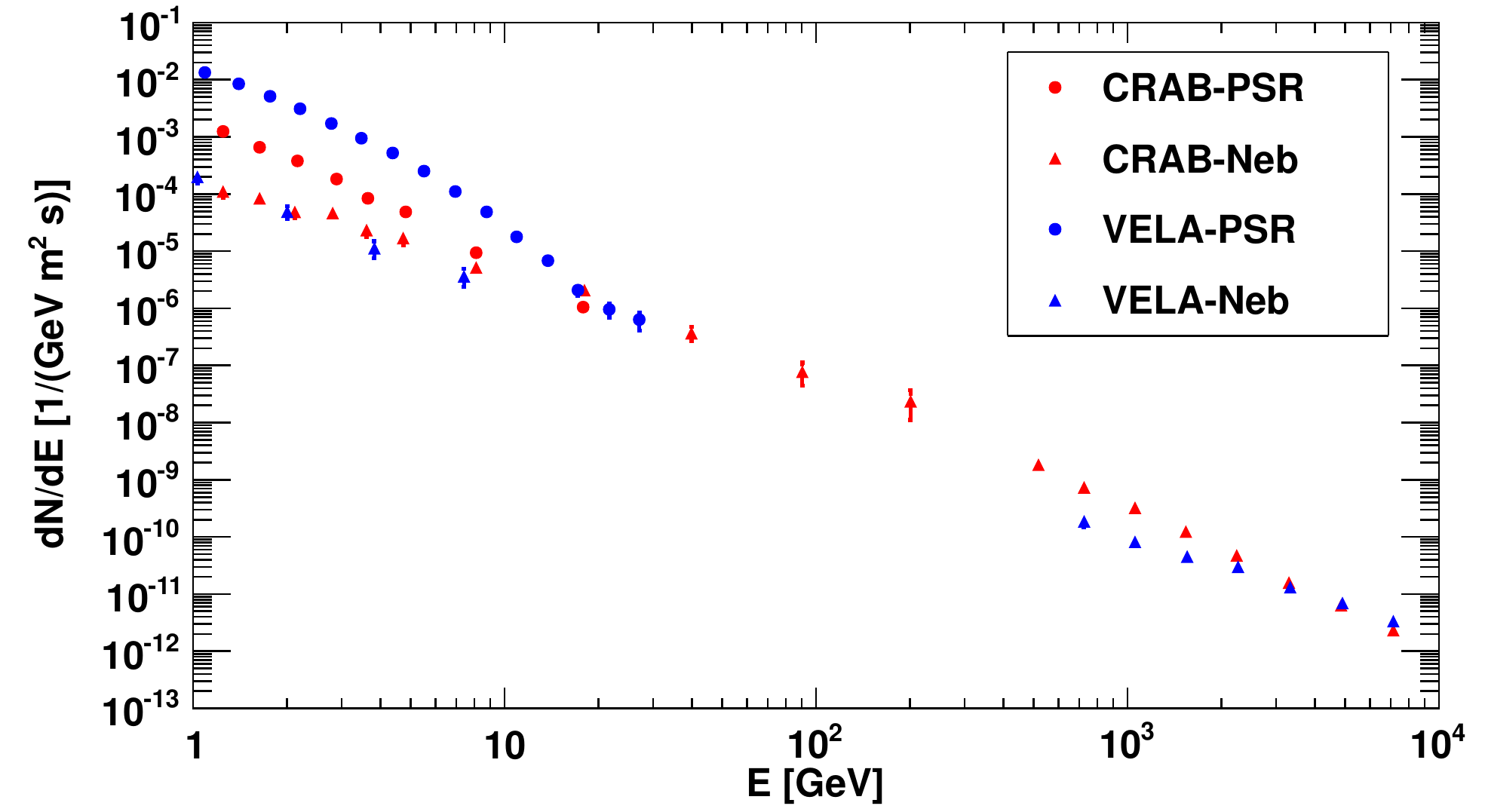}
  \caption{Gamma ray flux from Crab pulsar \cite{FermiCrab} in red circle and nebula \cite{HessCrab,FermiCrab} in red triangle and from Vela pulsar \cite{FermiVelaPSR} in blue circle and nebula \cite{HESS06,FermiVelaNeb} in blue triangle; the pulsar spectra are detected by Fermi collaboration, while the nebulae spectra is observed from Fermi at low energy and from HESS collaborations for energy above 500 GeV}
  \label{VelaCrabPSRNeb}
 \end{figure}

The pulsar and nebula spectra are usually described as function of energy with a power law and an exponential cutoff.
\begin{equation}\label{dNdE}
\frac{dN}{dE}\propto E^{-\alpha}e^{-\frac{E}{E_{cut}}}
\end{equation}
The Fermi experiment \cite{Abdo2010} found cutoff of less than 10 GeV for pulsars. Other experiments, like HESS \cite{HESS06}, have detected high energy gamma ray from nebula with cutoff of more than 1 TeV. For this reason, in this work we focused our attention on the nebula spectrum. The gamma ray spectrum of the Crab nebula \cite{HessCrab,FermiCrab} and its pulsar \cite{FermiCrab} is known, as well as that of Vela (nebula Vela-X \cite{HESS06,FermiVelaNeb} and pulsar \cite{FermiVelaPSR}) (see Fig. \ref{VelaCrabPSRNeb}).\\
Our key assumption is that these spectra represent the positron sources for a pulsar plus nebula source of about $10^{3}$ and $10^{4}$ years (Crab and Vela ages respectively).
We approximated the Crab spectra above 1 GeV with the eq. (\ref{dNdE}), spectral index $\alpha\sim2.2$ and an exponential energy cutoff at $E_{cut}\sim10$ TeV.\\
In our analysis we considered the same total energy output for photons and for particles. The luminosities of the Crab nebula and Vela-X are $4.6\cdot10^{35}$ erg/s and $1.1\cdot10^{33}$ erg/s respectively. These correspond at $\sim0.1\%$ and $\sim0.02\%$ of the pulsar spin-down luminosities.\\
We assumed the following power law relation between the gamma luminosity ($L_{\gamma}$) of Crab and Vela nebulae and their ages $\tau$.
\begin{equation}\label{LumEtà}
L_{\gamma}\propto\tau^{-\Gamma}
\end{equation}
For these two nebulae we found an index $\Gamma\sim2.5$. Using an injection spectrum for particles like:
\begin{equation}\label{QE}
Q(E)=Q_{0}E^{-\alpha}e^{-\frac{E}{E_{cut}}}
\end{equation}
we extracted the normalization factor ($Q_{0}$) deriving it from the photon total energy output.\\
At the birth of Vela-X we assumed the same spectrum of the Crab nebula. During its life we divided the spectrum in step of 1000 years. Each of these 10 steps (between the ages of the two objects) corresponds to a bunch of energy emitted at the end of the step. In this case the time between the end of the bunch and the present epoch is the diffusion time $t$. Using eq. (\ref{LumEtà}) we derived the normalization factor:
\begin{equation}\label{Q0}
Q_{0}=\frac{\int{L_{\gamma}(\tau)d\tau}}{\int{E^{1-\alpha}e^{-\frac{E}{E_{cut}}}dE}}.
\end{equation}
where the integral over time corresponds to the duration of the step (1000 years).

\section{The Model}
\begin{figure}[!t]
  \centering
  \includegraphics[width=0.45\textwidth]{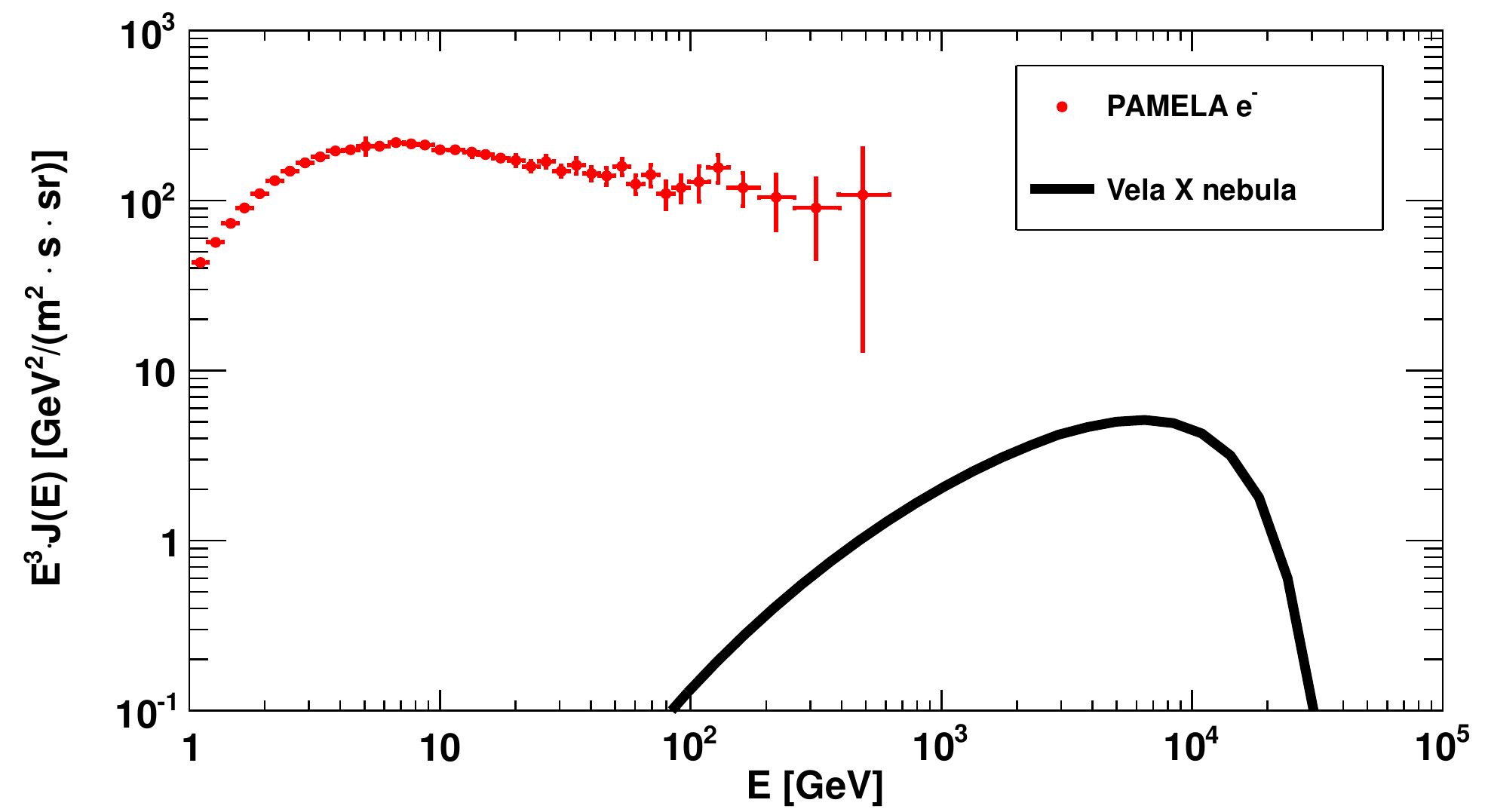}
   \caption{Vela nebula diffuse spectrum (in black) of positrons and electrons at the Solar System, in red dots the PAMELA \cite{Pamela2011} electron flux.}
   \label{AllVelaonly}
 \end{figure}

The time evolution of the energy density $N_{e}(\V x,E,t)$ for electrons and positrons is proposed in \cite{Kobayashi2004,Malyshev09}. The diffusion equation is:
\begin{eqnarray}\label{EqDiff}
  \frac{\partial N_{e}(\V x,E,t)}{\partial t} & = & Q(E)\nonumber\\
  & & +\vec{\nabla}\cdot\left[D(E)\vec{\nabla}N_{e}(\V x,E,t)\right]\\
  & & +\frac{\partial}{\partial E}\left[b(E)N_{e}(\V x,E,t)\right]\nonumber
\end{eqnarray}
where $E$ and $\V x$ are the observable energy of the particle and the distance between the pulsar and the Solar System. The terms $b(E)=b_0E^2$ and $D(E)$ represent the energy loss and the diffusion coefficient, both assumed spatial independent. $Q(E)$ is source term reported in eq. (\ref{QE}).\\
The analytic solution of this equation is reported in (\ref{flux}) where we defined the flux $J(E)=cN_{e}(E)/(4\pi)$ \cite{Malyshev09,Ginzburg64} and:
\begin{equation}\label{lambda}
\sigma_d(E,E_{0})=\int^{E_{0}}_{E}\frac{D(E')dE'}{b(E')}
\end{equation}
that contains the diffusion term. The differential intensity of the positrons or electrons injected from the source and diffuse in ISM is:
\begin{eqnarray}\label{flux}
J(\V x,E,t) & = & \frac{c}{4\pi}\frac{Q_0}{(4\pi\sigma_d)^{3/2}}E^{-\alpha}\left(1-b_0tE\right)^{\alpha-2}\\
            & & e^{-\frac{E}{E_{cut}(1-b_0tE)}}e^{-\frac{\V x^{2}}{4\sigma_d}}\nonumber
\end{eqnarray}
In our case we calculated the spectrum at the Solar System using the equation (\ref{flux}) for each bunch, we sum all the contributions and we reported the results in Fig. \ref{AllVelaonly}.\\
{\bf At 100 GeV the differential intensity of electrons (and positrons) from Vela-X is $\sim$0.1\% of the electron flux detected by PAMELA. Combining the informations from the Fig. \ref{AllVelaonly} and Fig. 6 reported in \cite{AMS02} we obtained that Vela-X contributes for about 1\% of the estimated positron flux at 100 GeV.}\\
The electron and positron spectrum from Crab is negligible because the time between the emission of the particles and the present time is less than the diffusion time ($\sim10^{6}$ years for a positron with $E=100$ GeV and distance between Crab and Solar System $\sim2$ kpc). We considered also the particle spectrum from the pulsars presented in the first Fermi-LAT catalog \cite{Abdo2010} using the model described above. Analysing the gamma luminosity and age of the pulsars we used the same power law of eq. (\ref{LumEtà}) with $\Gamma\sim0.4$ (we used the values of age and gamma luminosity reported in tables 1 and 4 of \cite{Abdo2010}). The total contribution of these pulsars is at low energy ($E<10$ GeV) and of one order of magnitude less than the Vela nebula contribution.\\
We also tried to interpret AMS-02 ratio with a pulsar with particular features like: $d=90$ pc, $\tau=9$ kyr, $\alpha=2.2$ and $E_{cut}=500$ GeV, but there no evidence of this object close to us.

\section{Conclusions}
We studied the possible contribution of primary particles from pulsars and nebulae for interpreting the positron fraction in cosmic rays. Under the assumption that the injection spectrum of electrons and positrons is the same observed in gamma rays, we compared the diffuse differential intensity with the experimental data. This contribution is hardly able to explain the observations of PAMELA and AMS-02. The reasons are that spectra from pulsars have a lower cutoff (less than 10 GeV), while the two nebulae contribute of about 1\% of the estimated positron flux at 100 GeV. The positron fraction of AMS-02 may be fitted by an hypothetical pulsar close to us, very young and with a high energy cut-off. The absence of observational evidences of such an object opens new hypothesis of the positron production in the Galaxy.

\vspace*{0.5cm}
\footnotesize{{\bf Acknowledgment:}{ This work is supported by Agenzia Spaziale Italiana under contract ASI-INFN I/002/13/0, Progetto AMS - Missione scientifica ed analisi dati.}}


\begin{thebibliography}{}
\bibitem{MS98} I.V. Moskalenko and A.W. Strong, ApJ, 493, 694, (1998).
\bibitem{AMS02} M. Aguilar et al., Phys. Rev. Lett., 110, 14, doi: 10.1103/PhysRevLett.110.141102, (2013).
\bibitem{Yin2013} P.F. Yin et al., ArXiv e-prints: 1304.4128, (2013).
\bibitem{Ibe2013} M. Ibe et al., ArXiv e-prints: 1305.0084, (2013).
\bibitem{Sturrock71} P.A. Sturrock, ApJ, 164, 529-556, (1971).
\bibitem{Abdo2010} A.A. Abdo et al., Astrophys.J.Suppl., 187, 460-494, doi: 10.1088/0067-0049/193/1/22, 10.1088/0067-0049/187/2/460, (2010).
\bibitem{HESS06} F. Aharonian et al., Astron.Astrophys., 448,  L43-L47, (2006).
\bibitem{HessCrab} F. Aharonian et al., Astron.Astrophys., 457, 899-915, doi: 10.1051/0004-6361:20065351, (2006).
\bibitem{FermiCrab} A.A. Abdo et al., ApJ, 708, 1254-1267, doi: 10.1088/0004-637X/708/2/1254, (2010).
\bibitem{FermiVelaNeb} The Fermi LAT Collaboration and Timing Consortium, ArXiv e-prints: 1002.4383, (2010).
\bibitem{FermiVelaPSR} A.A. Abdo et al., ApJ, 713, 154-165, doi: 10.1088/0004-637X/713/1/154, (2010).
\bibitem{Kobayashi2004} T. Kobayashi et al., ApJ, 601, 340-351, doi: 10.1086/380431, (2004).
\bibitem{Malyshev09} D. Malyshev, I. Cholis and J. Gelfand, ArXiv:0903.1310v3, (2009).
\bibitem{Ginzburg64} V.L. Ginzburg and S.I. Syrovatskii, Pergamon, Oxford, (1964).
\bibitem{Pamela2011} O. Adriani et al., Phys. Rev. Lett., doi: 10.1103/PhysRevLett.106.201101, (2011).
\bibitem{Vladimirov2011} A.E. Vladimirov et al., Computer Physics Communications, 182, 5, 1156-1161, doi: 10.1016/j.cpc.2011.01.017, (2011).
\bibitem{MS2004} A.W. Strong, I.V. Moskalenko and O. Reimer, ApJ, 613, 962-976, doi: 10.1086/423193, (2004).
\end{thebibliography}
\end{document}